\documentclass[twocolumn,showpacs,preprintnumbers,amsmath,amssymb]{revtex4}
\usepackage{amssymb}
\usepackage{amsfonts}
\usepackage{mathrsfs}
\usepackage{amsmath}
\usepackage{graphicx}
\usepackage{dcolumn}
\usepackage{bm}
\usepackage{hyperref}

\begin{document}
\title{Partially random phase attack to the practical two-way quantum key distribution system}
\author{Shi-Hai Sun, Ming Gao, Mu-Sheng Jiang, Chun-Yan Li, Lin-Mei Liang\footnote{Email:nmliang@nudt.edu.cn}}
\affiliation{Department of Physics, National University of Defense
Technology, Changsha 410073, People's Republic of China}
\begin{abstract}
Phase randomization is a very important assumption in the Bennett-Brassard 1984 quantum key distribution (QKD) system with a weak coherent source. Thus an active phase modulator is needed to randomize the phase of source. However, it is hard to check whether the phase of source is randomized totally or not in practical QKD systems. In this paper a partially random phase attack is proposed to exploit this imperfection. Our analysis shows that Eve can break the security of a two-way QKD system by using our attack, even if an active phase randomization is adopted by Alice. Furthermore, the numerical simulation shows that in some parameter regimes, our attack is immune to the one-deco-state method.
\end{abstract}


\maketitle
\section{\label{sec:Intro}Introduction}
Quantum key distribution (QKD) \cite{Bennett} admits two remote parties, known as Alice and Bob, to establish an unconditional secret key; even the eavesdropper (Eve) has unlimited power admitted by the quantum mechanics. The unconditional security of QKD has been proved in theory for both the ideal system \cite{Lo99,Shor00} and the practical system \cite{GLLP04,Inamori07} based on some assumptions. However, the practical QKD system is imperfect. Strictly speaking, any deviation between the standard security analysis and the practical QKD system can be exploited by Eve to attack the practical system \cite{Lydersen10,Gisin06,Zhao08,Makarov06, Fung07,Xu10,Sun11}. Therefore, in order to guarantee the unconditional security of the final key generated by the practical system, the legitimate parties must survey the practical QKD system carefully and close these loopholes.

In the standard security analysis for the Bennett-Brassard 1984 (BB84) QKD system with a weak coherent source (WCS), an important assumption is that the phase of source has been randomized totally. Thus, in the view of Eve, the state sent by Alice is a mixed state of all number states. However, in a practical QKD system, the phase information of source might be accessible to Eve \cite{Lo07}. In order to remove the phase randomization assumption from the standard security analysis, Lo and Preskill have proved the security of BB84 protocol using the WCS with nonrandom phase. But their proof is at the price of the secret key rate and the maximal security distance is very short \cite{Lo07}. Thus the best choice for the legitimate parties is to actively randomize the WCS phase, which can be implemented by modulating a totally random phase $\theta \in [0,2\pi]$ with a phase modulator \cite{Zhao07}. In fact, in most practical QKD systems \cite{Idq,Sasaki11,Hiskett06,GYS}, including the commercial system produced by Id Quantique \cite{Idq}, the legitimate parties assume the phase of source has been randomized totally, and thus they use the Gottensman-Lo-L\"{u}tkenhaus-Preskill (GLLP) formula \cite{GLLP04} but not the results of Ref.\cite{Lo07} to estimate the key rate. Specifically, the phase randomization assumption is the base of the decoy state method \cite{Hwang03,Lo05,Wang05,Ma05}, which is often used to defeat the photon number splitting (PNS) attack \cite{Huttner95,Brassard00}. Thus, in the QKD system with decoy-state method, only the GLLP formula can be used.

However, in practical situations, it is a hard task for the legitimate parties to check whether the phase of source has been randomized totally or not \cite{note1}. In the latter, we will show that even if an active \emph{phase modulator} is used by the legitimate parties to randomize the phase of source, Eve can change the range of random phase by using the imperfection of the phase modulator so that it is just partially randomized. Here \emph{partially random} means that the range of random phase modulated by Alice is smaller than $2\pi$. In other words, the random phase $\theta \in [0,\delta]$ and $\delta < 2\pi$. Furthermore, we note that in most of the practical system \cite{Idq,Sasaki11,Hiskett06,GYS}, no active setup is used to randomize the phase of source. We think there may be two reasons: (1) When Alice actively randomizes the source, an additional active setup is needed which will increase the complexity of the QKD system. (2) More importantly, within the best of our knowledge, until now there is not an effective attack strategy to exploit the phase information of WCS. In other words, Eve does not know how to spy the secret key, even if the source is not randomized or just partially randomized.

In this paper, we propose a partially random phase (PRP) attack to break the security of the practical two-way QKD system using WCS. Then a simple intercept-and-resend attack strategy and experimental arrangement within current technology are proposed to spy the secret key. Our analysis shows that the error rate induced by Eve can be lower than the tolerable threshold value of error rate, whereas the same range of error rate has been proved secure if the legitimate parties are unaware of our attack. Thus when our attack is taken into account, the secret key rate will be compromised. Specifically, the numerical simulations show that, in some parameter regime, our attack is immune to the one-decoy-state method \cite{Hwang03,Lo05,Wang05,Ma05} which is often used to defeat the PNS attack \cite{Huttner95,Brassard00}. Therefore, the legitimate parties should consider our attack carefully when they use the WCS to implement the BB84 protocol. However, note that we only claim that our attack is immune to the one-decoy-state method in some parameter regimes, but we do not claim that our attack is completely immune to the decoy-state-method. In fact, our attack can be defeated by the two-decoy-state (weak + vacuum) method \cite{Wang05,Ma05}, since the vacuum state is used to estimate the gain and error rate of the background.

We note that in Ref.\cite{Lo07}, Lo and Preskill have proposed a simple attack to exploit the nonrandom phase of source. However, our attack performs better than their attack at least in two aspects. First, in their attack Eve performs the positive operator valued measure (POVM) belonging to the photon number state space to distinguish the key bit, which can not be implemented within current technology. Our attack needs only a homodyne detector, which can be implemented within current technology. Second, in their attack Eve needs to know the exact phase of source, which corresponds to the case that the source is not randomized. But our attack is valid as long as the source is partially randomized. Therefore, the legitimate parties do not need to consider their attack, especially in practical situations, but they must consider our attack and monitor their system carefully.

The paper is organized as follows: In Sec. \ref{sec:attack} we introduce how to exploit the imperfection of a partially random phase to break the security of the cryptosystem. In Sec. \ref{sec:result} we introduce an intercept-and-resend attack with our PRP attack and then analyze the error rate induced by our attack. In Sec. \ref{sec:decoy} we show that our attack is immune to the one-decoy-state method which can be used to defeat the PNS attack. In Sec. \ref{sec:dis} we provide some discussion regarding our attack. Finally, in Sec. \ref{sec:summary} we give a brief summary of this paper.

\section{\label{sec:attack}PRP attack}
In this section, we first introduce the \emph{plug-and-play} QKD system briefly. Then we discuss how to exploit the partially random phase of source to spy the secret key.

\begin{figure}
\scalebox{1}{\includegraphics[width=\columnwidth]{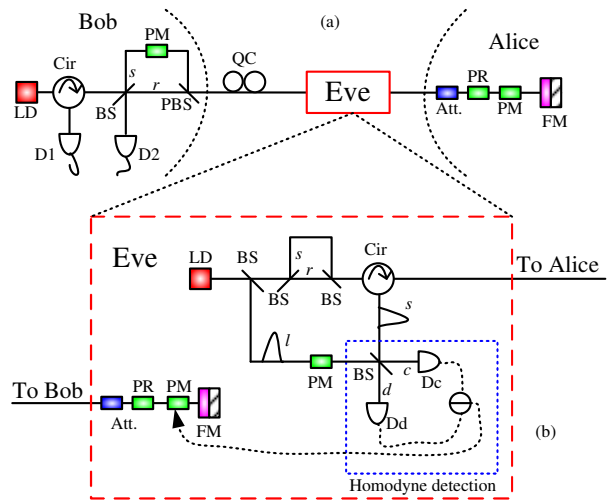}}
\caption{\label{fig:exp_arr}(Color online) A simple diagram of the \emph{Plug-and-Play} QKD system \cite{Muller97} and Eve's experimental arrangement. LD: laser diode; Cir: circulator; BS: 50/50 beam splitter; PBS: polarization beam splitter; Att.: attenuator; PM: phase modulator; PR: phase randomizer; FM: Faraday mirror; QC: quantum channel; D1 and D2: single photon detectors (SPDs); Dc and Dd: photodiode. Note that in the commercial system of Id Quantique, there is not PR \cite{Idq}. Part (a) shows the practical QKD system. Part (b) shows Eve's experimental arrangement. Eve intercepts the pulse from Bob and sends a faked pulse to Alice. When the faked pulse is modulated by Alice and returns to Eve, Eve measures it and modulates a phase on Bob's pulse according to her measurement results. Then she resends Bob's pulse to Bob. Note that in our attack Eve uses the homodyne detection but not SPD to detect Alice's information.}
\end{figure}

\subsection{Plug-and-Play System}
A simple diagram of a \emph{plug-and-play} QKD system \cite{Muller97} without Eve is shown in Fig.\ref{fig:exp_arr}(a). A strong pulse sent by Bob's laser will be divided into two parties by a 50:50 beam splitter (BS), noted as a signal pulse (\emph{s}) and reference pulse (\emph{r}). When the two pulses arrive at Alice's zone, Alice encodes her information on the signal pulse by modulating a phase $\phi_k^a=k\pi/2,k=0,1,2,3$. In order to ensure the global phase of pulse is totally random, Alice modulates a random phase $\theta \in[0,\delta]$ on both \emph{s} and \emph{r}. Note that $\theta$ should be random on each pulse of the laser, but it must be the same for both \emph{s} and \emph{r} of each pulse. Then the two pulses will be attenuated to a single photon level and sent back to Bob. Thus the two outgoing pulses from Alice's zone can be written as
\begin{equation}\label{state}
|\alpha e^{i(\theta+\theta')}\rangle_r |\alpha e^{ i(\phi_k^a+\theta+\theta')}\rangle_s,
\end{equation}
where $\theta'$ is the phase induced by the birefringence of fiber. $\alpha$ is real and $|\alpha|^2$ is the average photon number of \emph{s} and \emph{r}. Here we assume the Faraday mirror (FM) is perfect in the QKD system; thus the birefringence of fiber can be compensated successfully with a global phase $\theta'$. Otherwise Eve can obtain more information by combining with the passive FM attack proposed by our group \cite{Sun11}. Note that $\theta'$ is a fixed value and can be compensated by Eve, and thus we assume $\theta'=0$ in this paper. Therefore the practical state sent by Alice is given by
\begin{equation}\label{rho}
\rho=\int_{0}^{\delta} \frac{d\theta}{\delta} |\alpha e^{ i(\phi_k^a+\theta)}\rangle \langle \alpha e^{i(\phi_k^a+\theta)}|,
\end{equation}
where $\delta$ is range of random phase modulated by Alice. Here we assume $\theta$ follows the uniform distribution on $[0,\delta]$. For the three cases that the phase is nonrandomized, partially randomized, and totally randomized, $\delta=0$, $0<\delta<2\pi$, and $\delta=2\pi$, respectively. Note that in Eq.\eqref{rho} we consider only the signal pulse and ignore the reference pulse, since only the signal pulse is modulated by Alice.

Here we remark that if Alice does not actively randomize the phase of source or she thinks that Eve may know the phase of source exactly, she must use the method given by Ref. \cite{Lo07} to estimate the key rate. However, the security analysis of Ref. \cite{Lo07} can only be used for short-distance quantum cryptography. Thus, generally speaking, Alice should use an active setup to randomize the phase of source in the long-distance quantum cryptography. However, in the practical QKD system, it is a hard task for the legitimate parties to check whether the phase is really totally randomized or not. In fact, if Alice and Bob do not check carefully that the phase is truly random, then the range of random phase can be controlled by Eve so that it is just partially randomized ($\delta < 2\pi$). For example, in the two-way QKD system, since the phase modulator has a finite response time, Eve can control the practical phase modulated on the pulse by shifting the arrival time of the signal pulse to the rising or falling edge of the phase modulator. It is known as a phase remapping attack, which was proposed by Fung \emph{et al.} in theory \cite{Fung07} and then demonstrated by Xu \emph{et al.} in experiment \cite{Xu10}. Thus the practical phase modulated on the source will be lower than the expected phase of Alice, if Eve shifts the arrival time of the pulse. For instance, Alice wants to modulate a random phase $\theta'$, but the practical phase modulated on the source is $\theta<\theta'$. Therefore, in the practical QKD system, the range of random phase modulated by Alice can be controlled by Eve so that the pulse is just partially modulated ($\delta < 2\pi$).

\subsection{PRP Imperfection}
We have shown that the random phase can be controlled by Eve so that it is just partially modulated. In the following, we will show how Eve can exploit this imperfection to spy the secret key.

Since Alice admits the pulse into and out of her zone in the two-way system, it is easy for Eve to set her experimental arrangement to load our attack, which is shown in  Fig.\ref{fig:exp_arr}(b). The strong pulse sent by Eve's laser will be divided into three parts by two BS, signal pulse (\emph{s}), reference pulse (\emph{r}) and local pulse (\emph{l}). Then Eve sends \emph{s} and \emph{r} to Alice and keeps \emph{l} in her own hand. When the pulse is modulated by Alice and resent back to Eve, Eve modulates randomly \emph{l} with a phase $\phi_j^e=j\pi/2, j=0,1$. Then \emph{s} will interfere with \emph{l} at BS and be detected by two photodiodes (Dc and Dd).

Here Eve uses a homodyne detector to analyze Alice's information. The homodyne detection is a well-established quantitative method to measure the quadrature-amplitude operator of the signal field \cite{Yuen83,Leonhardt97} or implement the continuous variable quantum cryptography \cite{Symul07,Lodewyck07}. A simple illustration of the homodyne detection is shown in Fig.\ref{fig:exp_arr}(b).  The signal pulse (\emph{s}) and the local pulse (\emph{l}) interfere at a 50/50 beam splitter (BS). Then the two output modes of BS, \emph{c} and \emph{d}, will reach two photodiodes (Dc and Dd) respectively. The measured output signal of homodyne detection is determined by the difference of Dc and Dd, which is given by \cite{Hirano03}
\begin{equation}\label{i2x}
i \propto \sqrt{n_l}\langle a_s e^{-i\phi'}+a_s^+e^{i\phi'} \rangle\equiv \sqrt{2n_l} x,
\end{equation}
where $n_l$ is the average photon number of the local pulse, $a_s$ is the annihilation operator of signal pulse, $\phi'$ is the difference of phase between signal pulse and local pulse. Here we assume the local pulse is strong coherent state. $x$ is known as the normalized quadrature amplitude of the signal.  Generally speaking, $x$ takes a random value for each pulse due to the quantum fluctuations. Theoretically, the probability of $x$ is given by integrating the Wigner distribution \cite{Vogel89}, which is given by
\begin{equation}
P(x,\theta')=\int_{-\infty}^{\infty} W(x\cos\theta'-p\sin\theta', x\sin\theta'+p\cos\theta')dp,
\end{equation}
where $W(q,p)$ is the Wigner function in $p-q$ space.

It is easy to check that, when the signal pulse is  coherent state, the probability distribution of $x$ is given by \cite{Hirano03,Braunstein05}
\begin{equation}\label{P_theta}
P(x,\phi,\theta)=\sqrt{\frac{2}{\pi\kappa^2}}\exp[-2(x-\lambda \sqrt{\mu_s}\cos(\phi+\theta))^2/\kappa^2],
\end{equation}
where $\mu_s=|\alpha|^2$ is the average photon number of signal pulse, $\phi=\phi_k^a-\phi_j^e$ is the difference of phase modulated by Eve and Alice, $\theta\in[0,\delta]$ is the random phase modulated by Alice. $\lambda$ and $\kappa$ are two parameters that characterize the imperfection of homodyne detection \cite{Hirano03,note2}. When the homodyne is perfect, $\lambda=\kappa=1$. According to Eq.\ref{rho}, Eve has no priori information about $\theta$ excepting that $\theta\in[0,\delta]$, thus Eq.\eqref{P_theta} should be rewritten as
\begin{equation}\label{P_x}
P(x,\phi)=\int_{0}^{\delta}  \frac{d\theta}{\delta}\sqrt{\frac{2}{\pi\kappa^2}}\exp[-2(x-\lambda\sqrt{\mu_s}\cos(\phi+\theta))^2 /\kappa^2].
\end{equation}

\begin{figure}
\scalebox{1}{\includegraphics[width=\columnwidth]{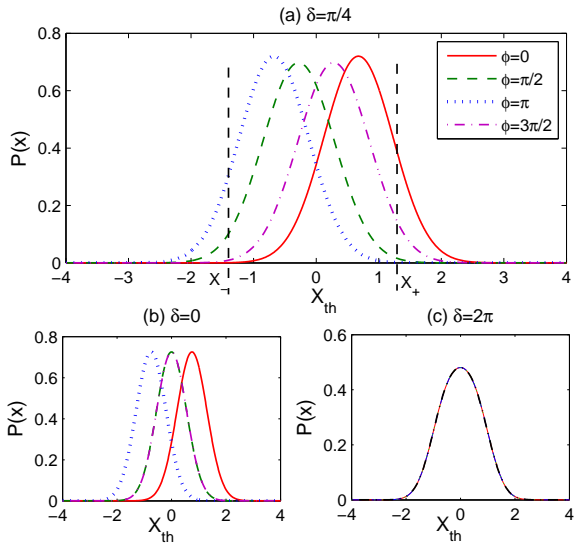}}
\caption{\label{fig:theo_dis}(Color online) The theoretical probability distribution of the quadrature amplitude when the total phase shifts are $0$, $\pi/2$, $\pi$ and $3\pi/2$. In the figure, we show the three case that the source has been unrandomized ($\delta=0$), partially randomized ($\delta=\pi/4$) and totally randomized ($\delta=2\pi$). Here we set $\lambda=0.75$ and $\kappa=1.1$ due to the experimental results of Ref.\cite{Hirano03}. $X_+$ and $X_-$ are two threshold values used to distinguish $0$ and $\pi$ from the set $\{0,\pi/2,\pi,3\pi/2\}$.}
\end{figure}

Fig.\ref{fig:theo_dis} shows the theoretical probability distribution of $x$ when the total phase shifts $\phi$ are $0$, $\pi/2$, $\pi$ and $3\pi/2$. It shows clearly that measured result of the homodyne detection can be used to distinguish $0$ and $\pi$ by setting a suitable threshold value. For example, Eve can set two threshold value $X_+ \geq 0$ and $X_- \leq 0$. When the measured quadrature amplitude $x \geq X_+$, Eve judge that $\phi=0$; when $x\leq X_-$, Eve judge that $\phi=\pi$. But when $X_- <x < X_+$, Eve can't give a judgement about the phase. Although Eve can't distinguish the four phases deterministically, it is easy to check that, the conditional probability that Eve obtain a valid outcome given that $\phi=0$ or $\phi=\pi$ can be much larger than that of $\phi=\pi/2$ and $\phi=3\pi/2$. Here, a valid outcome means the measured outcome $x$ satisfies $x\geq X_+$ or $x\leq X_-$.

Obviously, the smaller the probability Eve makes a wrong judgement, the larger $\mu_s$ is. Here \emph{wrong judgement} means Eve obtains $X_+$ (or $X_-$) but the state sent by Alice isn't $\phi=0$ (or $\phi=\pi$). In order to maximize Eve's information for a given system, we make two remarks about the optimal intensity of signal pulse ($\mu_s$) that is accessible for Eve.

\emph{Remark one}: Generally speaking, Alice will monitor the power of light coming into her zone, thus Eve should set the intensity of light sent to Alice carefully. However, note the fact that, Alice doesn't monitor the intensity of \emph{s} and \emph{r} respectively in most practical QKD systems. Thus Eve only needs to ensure the total of them entering Alice's zone, denoted as $n_a^i$, is unchanged, but no needs to keep the intensities of both \emph{s} and \emph{r} to be the same as their expected values respectively. In other words, Eve can change the proportion of average photon number between signal pulse and reference pulse (denoted as $n_s$ and $n_r$ respectively), but keeps $n_s+n_r=n_a^i$ constant. In the following, we let $n_s=\beta n_a^i$ and $n_r=(1-\beta)n_a^i$. Note that, if Alice  monitors the intensity of signal pulse and reference pulse at the same time, Eve can't change the proportion between $n_s$ and $n_r$. Then Eve must keep the intensity of $n_s$ and $n_r$ unchanged, which means $\beta=1/2$.

\emph{Remark two}: Note that only \emph{s} will be modulated by Alice, thus Eve should increase the coefficient $\beta$ to maximize her information. The maximal value of $\beta$ that can be set by Eve, denoted as $\beta_{max}$, depends on the way that Alice and Bob synchronize their clock in the practical QKD system. Generally speaking, there are two ways to synchronize the clock of Alice and Bob. One way is that Alice triggers her setups according to the arrival time of \emph{r}. Thus $(1-\beta_{max}) n_a^i$ is the minimal intensity of \emph{r} that can trigger Alice's setups, which depends on the efficiency of Alice's photodiode. The other way is that Alice synchronizes her clock with Bob according to another optical or electric signal, such as the wavelength division multiplexing \cite{Tanaka08}. In this case, Alice doesn't know whether \emph{r} arrives at her zone or not. Thus, Eve can block the reference pulse and only send \emph{s} to Alice, which means $\beta_{max}=1$. In this paper we assume that Alice and Bob use the second way to synchronize their clock, thus $n_s=n_a^i$. Therefore, the intensity of signal pulse outgoing Alice's zone is $\mu_s=\gamma n_s=\gamma n_a^i$, where $\gamma$ is the transmittance of Alice's attenuator.

\section{\label{sec:result}Intercept-and-Resend Attack with PRP}
We show that the partially random phase of source may leave a loophole for Eve to spy the secret key. Generally speaking, when an imperfection is found by Eve, she can combine all imperfections of the system and take advantage of all attack strategies to maximize her information of the key. However we only consider a simple intercept-and-resend attack in this paper, which clearly shows that the generated key will be compromised due to the partially random phase.

According to the analysis above, we can consider the following attack: Eve intercepts all the pulse from Alice and measures the quadrature amplitude ($x$) of the signal pulse with the experimental arrangement of Fig.\ref{fig:exp_arr}(b). In order to judge Alice's phase information, Eve modulates the local pulse randomly and equiprobably with one of the two phases $\phi_j^e=j\pi/2,j=0,1$. Then she sets two threshold value $X_+$ and $X_-$. Simply, we consider the symmetrical case that $X_+=-X_- \equiv X_{th}$ in this paper. When $x \geq X_{th}$, Eve judges that the phase modulated by Alice is the same as her, thus she resends a state with phase $\phi_j^e$ to Bob. When $x \leq -X_{th}$, Eve judges that the difference of phase modulated by Alice and her is $\pi$, thus she resends a state with phase $\phi_j^e+\pi$ to Bob. When $-X_{th} \leq x \leq X_{th}$, she blocks this pulse and resends a vacuum state to Bob. Note that these invalid judgements will not affect our attack, since the channel between Alice and Bob is lossy.

Obviously, sometimes, Eve may make a wrong judgement. However, it is easy to check that, when Eve uses the same basis as Alice, the probability that Eve obtains a valid outcome is much higher than that Eve uses the different basis with Alice. Here a valid outcome means that the measured outcome $x$ of Eve satisfies $x \geq X_{th}$ or $x \leq -X_{th}$. Generally speaking, the larger Eve sets $X_{th}$, the lower the error rate. But, at the same time, the probability that Eve obtains a valid outcome will decrease. Thus Eve should make a tradeoff between the error rate and efficiency when she loads her attack in a practical situation.

In order not to be discovered, Eve should ensure that the error rate induced by her attack is smaller than the tolerable threshold value of Alice and Bob. In the following, we analyze the error rate induced by Eve's attack and show that Eve can load our attack without being discovered by the legitimate parties. Without loss of generalization, we assume that the phase modulated by Alice is $\phi_0^a=0$. According to Eq.\eqref{P_x}, when Eve modulates the local pulse with a phase $\phi_0^e=0$, the probability that she obtains a valid outcome are given by
\begin{subequations}
\begin{equation}
P_0^+=\sqrt{\frac{2}{\pi\delta^2\kappa^2}} \int_{X_{th}}^\infty dx \int_0^\delta d\theta \exp[-2(x-\sqrt{\mu_s}\cos(\theta))^2/\kappa^2],
\end{equation}
\begin{equation}
P_0^-=\sqrt{\frac{2}{\pi\delta^2\kappa^2}} \int_{-\infty}^{-X_{th}}dx \int_0^\delta d\theta \exp[-2(x-\sqrt{\mu_s}\cos(\theta))^2/\kappa^2].
\end{equation}
\end{subequations}
Obviously, when Eve obtains $x \geq X_{th}$, she will not induce any error. But when Eve obtains $x \leq -X_{th}$, she will induce an error event with a probability up to 1. At the same time, Eve may modulate the local pulse with a phase $\phi_1^e=\pi/2$, then the probability that she obtains a valid outcome are given by
\begin{subequations}
\begin{equation}
P_{\pi/2}^+=\sqrt{\frac{2}{\pi\delta^2\kappa^2}} \int_{X_{th}}^\infty dx \int_0^\delta d\theta \exp[-2(x+\sqrt{\mu_s}\sin(\theta))^2/\kappa^2],
\end{equation}
\begin{equation}
P_{\pi/2}^-=\sqrt{\frac{2}{\pi\delta^2\kappa^2}} \int_{-\infty}^{-X_{th}}dx \int_0^\delta d\theta \exp[-2(x+\sqrt{\mu_s}\sin(\theta))^2/\kappa^2].
\end{equation}
\end{subequations}
Obviously, she will induce an error event with a probability 1/2 for this case, no matter which outcome is obtained by her. Therefore, the error rate induced by Eve can be written as
\begin{equation}\label{error_rate}
e=\frac{P_0^-+[P_{\pi/2}^++P_{\pi/2}^-]/2} {P_0^+ +P_0^- +P_{\pi/2}^+ +P_{\pi/2}^-}.
\end{equation}
Here we assume Eve modulates randomly $0$ or $\pi/2$ with equal probability. Since the four states sent by Alice are symmetrical, the total error rate induced by Eve's attack is the same as Eq.\eqref{error_rate}. At the same time, the probability that Eve obtains a valid outcome is given by:
\begin{equation}\label{post}
P_{post}=(P_0^+ + P_0^- + P_{\pi/2}^+ + P_{\pi/2}^-)/2.
\end{equation}

\begin{figure}
\scalebox{1}{\includegraphics[width=\columnwidth]{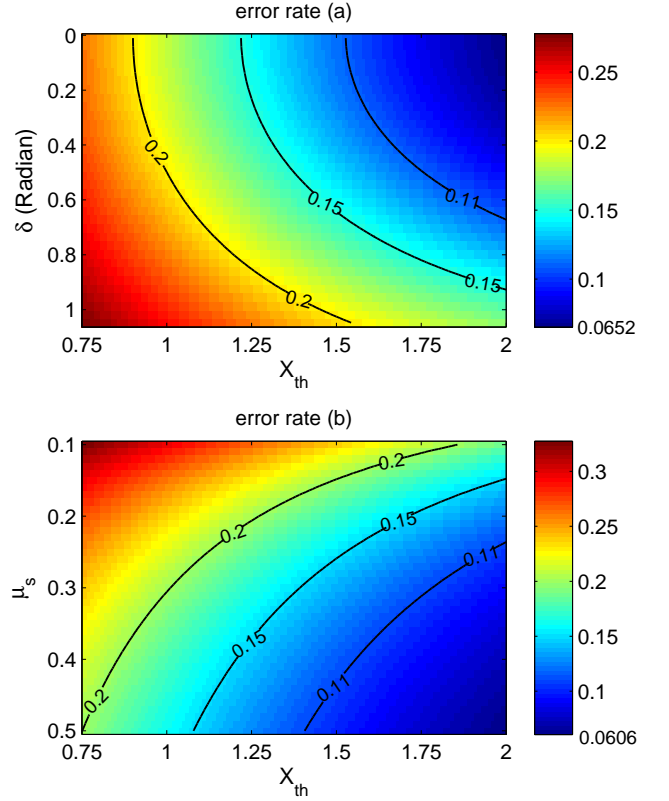}}
\caption{\label{fig:error_rate}(Color online) The error rate induced by Eve. In the simulations, we set $\lambda=0.75$ and $\kappa=1.1$ due to the experimental results of Ref.\cite{Hirano03}. Part(a) shows the error rate changes with $X_{th}$ and $\delta$ for given $\mu_s=0.3$. Part(b) shows the error rate changes with $X_{th}$ and $\mu_s$ for given $\delta=\pi/6$. In the figure, we also draw the contour line for $e=11\%,15\%, 20\%$.}
\end{figure}

The error rate induced by our attack is shown in Fig.\ref{fig:error_rate}. It shows clearly that when the phase of source is just partially random, the generated key will be compromised. In part(a) we show that the error rate changes with $X_{th}$ and $\delta$. As expected, for given $\delta$, the higher the threshold value $X_{th}$ is set by Eve, the lower the error rate will be. For example, when $\delta=\pi/6$ and $\mu_s=0.3$, the error rate is 9.21\% and 13.79\% for $X_{th}=2$ and $X_{th}=1.5$ respectively. At the same time, the smaller the range of partially random phase $\delta$ is, the lower the error rate will be. For example, when Eve sets $X_{th}=2$, the error rate is 8.01\%, 9.21\% and 12.65\% for $\delta=\pi/8$, $\delta=\pi/6$ and $\delta=\pi/4$ respectively. According to Eq.\ref{P_x}, the error rate induced by Eve will depend on the intensity of signal pulse $\mu_s$, which is shown in Fig.\ref{fig:error_rate} (part(b)). For example, when Eve sets $X_{th}=2$ and $\delta=\pi/6$, the error rate will be 6.06\%, 9.21\% and 18.65\% for $\mu_s=0.5,0.3,0.1$ respectively. In other words, if Eve wants to keep the error rate is smaller than 20\%, she only needs to set $X_{th}=1.02$ for $\mu_s=0.3$, but $X_{th}$ will increase to 1.86 for $\mu_s=0.1$

According to the analysis above, we know that Eve can reduce the error rate induced by her attack by increasing the threshold value $X_{th}$. However, the higher $X_{th}$ is, the lower Eve obtains an unambiguous result that $x \geq X_{th}$ or $x \leq -X_{th}$. Although Eve can send vacuum state to Bob when she obtains the ambiguous result that $-X_{th} < x < X_{th}$, she should ensure that the expected count rate of Bob is unchanged. In other words, Eve should ensure that the equation that $P_{post}\eta_{Bob}\mu_E =\mu_s\eta_c\eta_{Bob}$ holds. Here $\eta_{Bob}$ is the transmittance of Bob's setups. $\eta_c=10^{-al/10}$ is the transmittance of channel. $\mu_E$ is the intensity of pulse sent to Bob by Eve. Here we assume that Eve can send a strong pulse to Bob to compensate $\eta_{Bob}$, which means $\mu_E=1/\eta_{Bob}$.  Therefore, the maximal value of $X_{th}$ is determined by the transmittance of channel between Alice and Bob. The equivalent length of channel for a given $X_{th}$ is given by
\begin{equation}
l=-\frac{10}{a}\log_{10}[\min\{1,\mu_E P_{post}/\mu_s\}],
\end{equation}
where we assume the channel is fiber and $a=0.21dB/km$ is the typical loss of fiber and $P_{post}$ is given by Eq.\ref{post}.

\begin{figure}
\scalebox{1}{\includegraphics[width=\columnwidth]{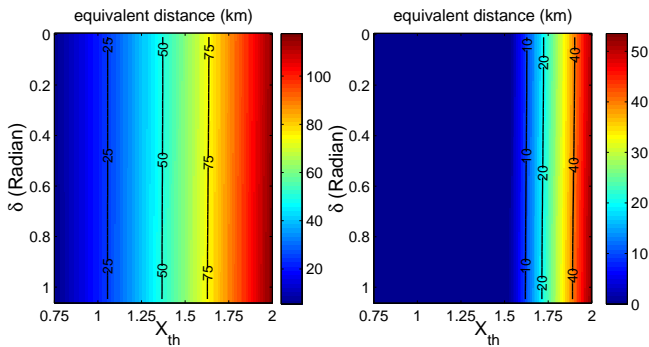}}
\caption{\label{fig:equivalent_length}(Color online)The equivalent length of channel that Eve can load her attack successfully for a given $X_{th}$. Here we assume the channel is fiber and $\mu_s=0.3$,  In the simulations, we set $\eta_{Bob}=0.045$, $\lambda=0.75$ and $\kappa=1.1$ due to the experimental result of \cite{GYS,Hirano03}. Part(a) shows the equivalent distance when Eve sends a single photon state to Bob, which means $\mu_E=1$ and the detection efficiency of Bob $\eta_{Bob}$ is not compensated. Part(b) shows the case that $\mu_E=1/\eta_{Bob}$, which means that Eve compensates totally Bob's detection efficiency $\eta_{Bob}$ by sending a strong pulse to Bob.}
\end{figure}

Fig.\ref{fig:equivalent_length} shows the equivalent length of channel changes with $X_{th}$ and $\delta$. It shows clearly that our attack is valid even the channel distance of Alice and Bob is very short. It can be explained as the maximal $X_{th}$ that Eve can set for a given length of channel. Note that the error rate induced by Eve is determined by $X_{th}$. Thus it also can be explained as the minimal error rate induced by our attack for a given length of channel. For example, if the length of channel between Alice and Bob is 50 km, the maximal $X_{th}$ that Eve can set is 1.97 when $\delta=\pi/6$ and $\mu_s=0.3$, thus the minimal error rate induced by Eve is 9.36\%.

\section{\label{sec:decoy}One Decoy state method}
It is well known that, in the practical QKD systems with the WCS, the decoy state method should be used to beat the PNS attack. In this section, we will show that, if Alice and Bob use the one decoy state method to estimate the final secret key rate, Eve still can spy the secret key using our attack in some parameter regime. Although the one decoy state is not optimal for Alice and Bob, it is still adopted in some experimental systems \cite{Zhao06,Hiskett06}.

In the one decoy state method, Alice will send two kinds of pulses with different intensities to Bob, the signal state and the decoy state, whose average photon number are denoted as $\mu$ and $\nu$ ($\mu>\nu>0$) respectively.
Then they estimate the lower bound for the yield of single photon state, $Y_1^L$, and the upper bound for error rate of single photon state, $e_1^U$, which are given by the Eq.(41) of Ref.\cite{Ma05}:
\begin{subequations}\label{Y1_e1}
\begin{equation}\label{Y1}
Y_1^L=\frac{\mu}{\mu\nu-\nu^2}[Q_\nu e^\nu -Q_\mu e^\mu \frac{\nu^2}{\mu^2}],
\end{equation}
\begin{equation}\label{e1}
e_1^U=\frac{E_\nu Q_\nu e^\nu}{Y_1^L \nu},
\end{equation}
\end{subequations}
where $Q_\mu$ and $Q_\nu$ are the gain of the signal state and the decoy state, $E_\mu$ is the error rate of the decoy state, $e_0=1/2$ is the error rate of background. Note that Alice and Bob will think that the phase of source has been randomized totally (in fact it is just partially randomized), thus they will use the GLLP formula to estimate the key rate, which is given by \cite{GLLP04}:
\begin{equation}\label{key_rate}
R\geq q\{-Q_\mu f(E_\mu) H_2(E_\mu)+ Q_1^L[1-H_2(e_1^U)]\},
\end{equation}
where $q=1/2$ with the standard BB84 protocol. $f(e)=1.22$ is the bidirectional error correction efficiency. $H_2(x)=-x\log_2(x)-(1-x)\log_2(1-x)$ is the binary Shannon information entropy. $Q_1^L=\mu e^{-\mu}Y_1^L$ is the gain of single photon state, $E_\mu$ is the error rate of the signal state.

Note that, in the decoy state method, Eve can't distinguish the signal state and the decoy state. Thus we assume Eve sends a single photon state to Bob, when she obtains a valid measurement outcome. Thus the gain and error rate of Bob for the signal state and the decoy state are given by
\begin{equation}\label{gain}
\begin{matrix}
Q_\mu=\eta_{Bob}Q'_\mu+(1-\eta_{Bob})Y_0,\\
E_\mu Q_\mu=\eta_{Bob}Q'_\mu E'_\mu+(1-\eta_{Bob})Y_0e_0,\\
Q_\nu=\eta_{Bob}Q'_\nu+(1-\eta_{Bob})Y_0,\\
E_\nu Q_\nu=\eta_{Bob}Q'_\nu E'_\nu+(1-\eta_{Bob})Y_0e_0,\\
\end{matrix}
\end{equation}
where $Y_0$ is the dark count of Bob's single photon detector, $e_0=0.5$ is the error rate of dark count. Under our attack, $E'_\mu$ and $E'_\nu$ are given by Eq.\eqref{error_rate}, $Q'_\mu$ and $Q'_\nu$ are given by Eq.\eqref{post}.

\begin{figure}
\scalebox{1}{\includegraphics[width=\columnwidth]{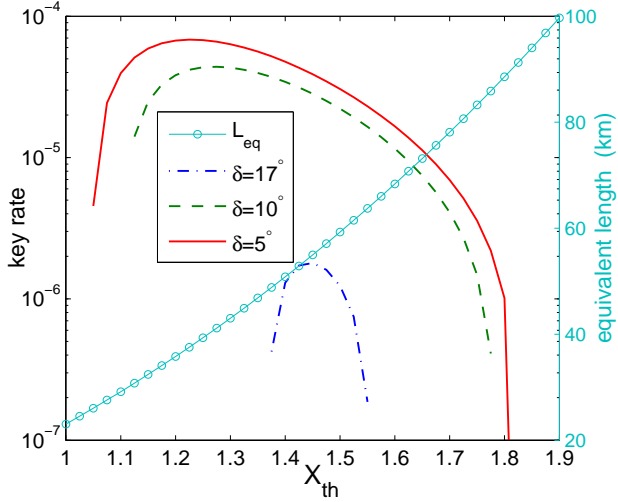}}
\caption{\label{fig:key_rate}(Color online) The key rate between Alice and Bob at various threshold value $X_{th}$. Here the legitimate parties use the one decoy state method to estimate the key rate. Here we set $\mu=0.48$ and $\nu=0.05$ according to the decoy state theory \cite{Ma05} and the experimental parameters of GYS \cite{GYS}, which are laser $\lambda=1550$ nm at 2 MHz, dark count rate $Y_0=1.7\times10^{-6}$, fiber loss 0.21 dB/km, Bob's quantum efficiency $\eta_{Bob}=0.045$. In the simulation, we assume the homodyne detector of Eve is perfect, which means $\lambda=\kappa=1$. Strictly speaking, the equivalent length will change with $\delta$, but the difference is much small. Thus we just draw the equivalent length for $\delta=17^\circ$ in the figure.}
\end{figure}

Substituting these parameters into Eq.\eqref{key_rate}, it is easy to estimate the key rate under our attack, which is shown in Fig.\ref{fig:key_rate}. Here we also show the equivalent length of channel between Alice and Bob, which is given by
\begin{equation}
L_{eq}=-\frac{10}{a}\log_{10}[\min\{1,\frac{Q_{\mu}}{\mu\eta_{Bob}}\}],
\end{equation}
where $Q_{\mu}$ is the gain of signal state. Here the equivalent length is used to ensure the gain of signal state under our attack is same as Bob's expectancy.

The numerical simulations show clearly that, Eve can ensure the key rate between Alice and Bob is still positive by setting a suitable threshold value.
For example, when $\delta=17^\circ$, the key rate is positive if Eve sets $X_{th}> 1.37$, but in fact no secret key can be generated in this range, since Eve's intercept-and-resend attack is an entanglement breaking channel \cite{Fung07,Curty04}. Fig.\ref{fig:key_rate} also shows clearly that our attack can be implemented successfully even the distance of channel between Alice and Bob is short. For example, if Eve sets $X_{th}=1.4$, the key rate will be positive and the equivalent length of channel is about 50.83 km.

\begin{figure}
\scalebox{1}{\includegraphics[width=\columnwidth]{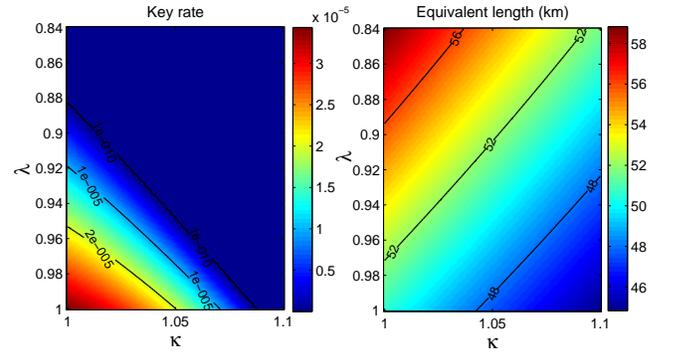}}
\caption{\label{fig:key_rate_imper}(Color online) The key rate between Alice and Bob changes with the parameters that characterize the imperfection of homodyne detection. In the simulation, we also use the experimental parameters of GYS (see Fig.\ref{fig:key_rate}) and set $\mu=0.48$ and $\nu=0.05$. Here we assume the partially randomized phase is $\delta=10^\circ$ and Eve sets $X_{th}=1.4$. The point that $\lambda=\kappa=1$ represents a perfect homodyne detector. In the figure, we also draw the equivalent length of channel when Eve sets $X_{th}=1.4$.}
\end{figure}

Note that, in the simulation of Fig.\ref{fig:key_rate}, we assume the homodyne detector of Eve is perfect. Although Eve can make a perfect homodyne detector in theory, a practical Eve is still imperfect. In Fig.\ref{fig:key_rate_imper}, we show the key rate changes with the two parameters $\lambda$ and $\kappa$ which are used to characterize the imperfections of Eve's homodyne detector. It shows clearly that our attack is still valid even the homodyne detector of Eve is imperfect. In Fig.\ref{fig:key_rate_imper}, we also draw the equivalent length of channel with the same parameters as that of key rate. It shows clearly that even the length of channel between Alice and Bob is short, our attack is still valid.

\begin{figure}
\scalebox{1}{\includegraphics[width=\columnwidth]{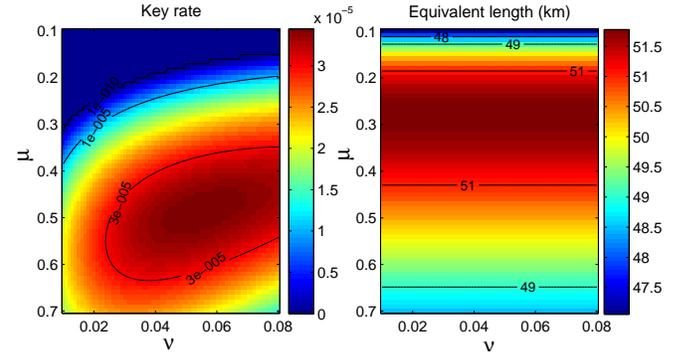}}
\caption{\label{fig:key_rate_mu_nu}(Color online)The key rate changes with the intensity of signal state and decoy state. In the simulation, we assume the homodyne detector is perfect and Eve sets $X_{th}=1.4$. Here the partially randomized is set as $\delta=10^\circ$. The equivalent length is also drawn in the figure to show that our attack is valid even in the short-distance QKD system.}
\end{figure}

In Fig.\ref{fig:key_rate_mu_nu}, we also show the key rate changes with the intensity of signal state and decoy state. Generally speaking, for a given practical QKD system, the legitimate parties will optimize the intensity of signal state and decoy state ($\mu$ and $\nu$) to maximize the key rate. For example, in the experimental parameters of Gobby-Yuan-Shield (GYS), the optimal $\mu$ and $\nu$ are 0.48 and 0.036 when the length of channel between Alice and Bob is 50km. However, under these parameters, the final key rate is still positive under our attack, but in fact no secret key can be generated. In other words, when our attack is taken into account, the estimated optimal parameters that are unaware of our attack may be not secret. Thus, in a practical system, the legitimate parties must set their experimental parameters carefully.

Here we remark that, although our attack is immune to the one decoy state method, it can be defeated by the vacuum+weak decoy state method, in which Alice sends three kinds of pulses to Bob, the signal state, the decoy state and the vacuum state. The main reason is that Eve may obtain valid outcome with a high probability, even Alice sends a vacuum state. Thus the gain of vacuum state will be much larger than the dark count of single photon detector (SPD), and then can be found by Alice and Bob.

\section{\label{sec:dis}Discussion}
We have shown that our attack can be used by Eve to break the security of a practical QKD system with WCS, even the one decoy state method is used by Alice and Bob. In the end of this paper, we give some discussions about our attack.

First, it has shown that our attack can beat the BB84 QKD system with WCS and SPD. However, it is invalid for the continuous variable QKD (CV-QKD) scheme. In the CV-QKD, Bob also uses the homodyne detection to measure the signal pulse, thus he can reconstruct the probability distribution of signal pulse and then discover the existence of Eve. Furthermore, due to the same reason as the CV-QKD, our attack is also invalid for the BB84 QKD system with pulse homodyne detection, which is proposed and demonstrated by Hirano \cite{Hirano03} \emph{et al}.

Second, in this paper, we assume that Eve only changes the random phase of source which is a global phase, but not remaps the bit phase which is a relative phase. In other words, in the analysis above, we assume the states sent by Alice are still the standard BB84 states. Strictly speaking, Eve also can change the bit phase. Then Eve can combine our attack with the phase remapping attack \cite{Fung07,Xu10} to maximize her information.

Third, in high-speed QKD system, Eve needs a high-speed homodyne detector to detect the signal pulse. It seems an experimental challenge to implement our attack. In fact, the speed of some commercial homodyne detectors produced by Picometrix can reach 40GHz \cite{picometrix}. Thus our attack is valid even for the high-speed QKD system. Further, a well-stable interferometer is needed to ensure that the local pulse and signal pulse can interfere at the BS. Thus Eve must compensate the phase shift of fiber induced by the environment. In fact, the feedback method developed in the GHz QKD system \cite{Sasaki11} can also be used in our attack to ensure the interferometer of Eve is very stable.

\section{\label{sec:summary}Summary}
For the BB84 QKD protocol, phase randomization is a very important assumption in the standard security analysis. However, it is a hard task to check whether the phase of source is randomized totally or not in the practical situation. In fact, Eve can control the range of random phase so that it is just partially randomized.

In this paper, we proposed a partially random phase attack to spy the secret key. Our results show that if the phase of source is partially randomized, the error rate induced by our attack can be lower than the tolerable threshold value of error rate whereas the same range of error rate has been proved secure if the legitimate parties are unaware of our attack. Thus the secret key generated by a practical QKD system will be compromised. Specially, the numerical simulation shows that our attack is immune to the one decoy state method in some parameter regime. Therefore, the legitimate parties should consider our attack carefully, when they use the WCS to implement the BB84 protocol.

\section{ACKNOWLEDGEMENT}
This work is supported by the National Natural Science Foundation of China, Grant No. 61072071. S.H. Sun is supported by the Hunan Provincial Innovation Foundation for Postgraduates, Grant No. CX2010B007, and the Fund of Innovation, Graduate School of NUDT, Grant No. B100203.



\begin{thebibliography}{100} 
\bibliographystyle{unsrt}

\bibitem{Bennett} C. H. Bennett and G. Brassard, \emph{Proceedings
of the IEEE International Conference on Computers, Systems and
Signal Processing, Bangalore, India} (IEEE, New York, 1984), pp. 175-179.

\bibitem{Lo99} H. -K. Lo, and H. F. Chau, Science, \textbf{283}, 2050 (1999).

\bibitem{Shor00} P. W. Shor, and J. Preskill, Phys. Rev. Lett. \textbf{85}, 441 (2000).

\bibitem{GLLP04} D. Gottesman, H. -K. Lo, N. L\"{u}tkenhaus, and J. Preskill, Quantum Inf. Comput. \textbf{4(5)}, 325 (2004).

\bibitem{Inamori07} H. Inamori, N. L\"{u}tkenhaus, and D. Mayers, Eur. Phys. J. D, \textbf{41}, 599 (2007).

\bibitem{Lydersen10} L. Lydersen, C. Wiechers, C. Wittmann, D. Elser, J. Skaar, and V. Makarov, Nat. Photonics, \textbf{4}, 686 (2010).

\bibitem{Gisin06} N. Gisin, S. Fasel, B. Kraus, H. Zbinden, and G. Ribordy, Phys. Rev. A \textbf{73}, 022320 (2006).

\bibitem{Zhao08} Y. Zhao, C. H. Fred Fuang, B. Qi, C. Chen, and H.-K. Lo, Phys. Rev. A \textbf{78}, 042333 (2008).

\bibitem{Makarov06} V. Makarov, A. Anisimov, and J. Skaar, Phys. Rev. A \textbf{74}, 022313 (2006).

\bibitem{Fung07} C. H. Fred Fung, B. Qi, K. Tamaki, and H. -K. Lo, Phys. Rev. A \textbf{75}, 032314 (2007).

\bibitem{Xu10} F. -H. Xu, B. Qi, and H. -K. Lo, New J. Phys. \textbf{12}, 113026 (2010).

\bibitem{Sun11} S. -H. Sun, M. -S. Jiang, and L. -M. Liang, Phys. Rev. A \textbf{83}, 062331 (2011).

\bibitem{Lo07} H. -K. Lo, and J. Preskill, Quantum Inf. Comput. \textbf{5(6)} 431 (2007).

\bibitem{Zhao07} Y. Zhao, B. Qi, and H. -K. Lo, Appl. Phys. Lett. \textbf{90}, 044106 (2007).

\bibitem{Idq} http://www.idquantique.com/scientific-instrumentation/clavis2-qkd-platform.html

\bibitem{Sasaki11} M. Sasaki, M. Fujiwara, H. Ishizuka, W. Klaus, K. Wakui, M. Takeoka, S. Miki, T. Yamashita, Z. Wang, A. Tanaka, \emph{et al.}, Opt. Express \textbf{19}, 10387 (2011).

\bibitem{Hiskett06} P. A. Hiskett, D. Rosenberg, C. G. Peterson, R. J. Hughes, S. Nam, A. E. Lita, A. J. Miller, and J. E. Nordholt, New J. Phys. \textbf{8} 193 (2006).

\bibitem{GYS} C. Gobby, Z. L. Yuan, and A. J. Shields, Appl. Phys. Lett. \textbf{84}, 3762 (2004).

\bibitem{Hwang03} W. -Y. Hwang, Phys. Rev. Lett. \textbf{91}, 057901 (2003).

\bibitem{Lo05} H. -K. Lo, X. -F. Ma, and K. Chen, Phys. Rev. Lett. \textbf{94}, 230504 (2005).

\bibitem{Wang05} X. -B. Wang, Phys. Rev. Lett. 94, 230503 (2005).

\bibitem{Ma05} X. -F. Ma, B. Qi, Y. Zhao, and H. -K. Lo, Phys. Rev. A \textbf{72}, 012326 (2005).

\bibitem{Huttner95} B. Huttner, N. Imoto, N. Gisin, and T. Mor, Phys. Rev. A \textbf{51}, 1863 (1995).

\bibitem{Brassard00} G. Brassard, N. L\"{u}tkenhaus, T. Mor, and B.C. Sander, Phys. Rev. Lett. 85, 1330 (2000).

\bibitem{note1} Here we means that it is hard for Alice to monitor the degree of random phase using current QKD system. Of course, if she adds additional optical and electrical setups, she can do this easily. For example, she can establish one interferometer to measure the visibility of pulses. If the visibility is zero, the phase of pulse is randomized totally. Otherwise, it is just partially randomized. Therefore, if our attack is taken into account, Alice must redesign her system to monitor carefully the random phase.

\bibitem{Muller97} A. Muller, T. Herzog, B. Huttner, W. Tittel, H. Zbinden, and N. Gisin, Apple. Phys. Lett. \textbf{70}, 793 (1997)

\bibitem{Yuen83} H. P. Yuen, and V. W. S. Chan, Opt. Lett. \textbf{8}, 177 (1983).

\bibitem{Leonhardt97} U. Leonhardt, \emph{Measuring the Quantum State of Light} (Cambridge, University Press, Cambridge, 1997).

\bibitem{Symul07} T. Symul, D. J. Alton, S. M. Assad, \emph{et al.}, Phys. Rev. A \textbf{76}. 030303(R) (2007).

\bibitem{Lodewyck07} J. Lodewyck, T. Debuisschert, \emph{et al.}, Phys. Rev. Lett. \textbf{98}, 030503 (2007).

\bibitem{Hirano03} T. Hirano, H. Yamanaka, M. Ashikaga, T. Konishi and R. Namiki, Phys. Rev. A \textbf{68}, 042331 (2003).

\bibitem{Vogel89} K. Vogel and H. Risken, Phys. Rev. A \textbf{40}, 2847 (1989).

\bibitem{Braunstein05} S. Braunstein and P. van Loock, Rev. Mod. Phys. \textbf{77}, 513 (2005).

\bibitem{note2} Here $\lambda=\sqrt{T\eta_{PD}}V$, where $T$ is the loss of optical setups, $\eta_{PD}$ is the quantum efficiency (QE) of photodiode, $V$ is the visibility of interference. Thus the crucial imperfection of homodyne detection is the QE of photodiode. Although the QE is just 0.84 in the experiment of Ref.\cite{Hirano03}, the QE of Hamamatsu's new product can approach 0.9 (see the website of Hamamatsu: http://jp.hamamatsu.com/products/sensor-ssd/pd128/pd129/pd130/G8370-81/index$_{-}$en.html). Furthermore, $\kappa$ is determined by the excess noise of the signal and the phase error. But, as the best of our knowledge, we only find one experimental result that $\kappa=1.1$ which is given by Ref.\cite{Hirano03}. However, it does not matter, theoretically, Eve can make a perfect homodyne detection with $\lambda=\kappa=1$.

\bibitem{Tanaka08} A. Tanaka, M. Fujiware, S. W. Nam, Y. Nambu, S. Takahashi, W. Maeda, K. I. Yoshino, S. Miki, B. Baek, Z. Wang, \emph{et al.}, Opt. Express \textbf{16}, 11354 (2008).

\bibitem{Zhao06} Y. Zhao, B. Qi, X.-F. Ma, H.-K. Lo, and L. Qian, Phys. Rev. Lett. \textbf{96}, 070502(2006).

\bibitem{Curty04} M. Curty, M. Lewenstein, and N. L\"{u}tkenhaus, Phys. Rev. Lett. \textbf{92}, 217903 (2004).
\bibitem{picometrix} http://www.picometrix.com/pico-products/hsor-datatables.asp

\end{thebibliography}
\end{document}